# Evaluation of flashing stimuli shape and colour heterogeneity using a P300 brain-computer interface speller

Álvaro Fernández-Rodríguez[1], Francisco Velasco-Álvarez, María Teresa Medina-Juliá and Ricardo Ron-Angevin.

E-mail: afernandezrguez, fvelasco, maytemed, rron {@uma.es}

Departamento de Tecnología Electrónica, Universidad de Málaga, Malaga, Spain.

**Abstract**

**Objective:** Previous works using a visual P300-based speller have reported an improvement modifying the shape or colour of the presented stimulus. However, the effects of both blended factors have not been yet studied. Thus, the aim of the present work was to study both factors and assess the interaction between them.

**Method:** Fifteen naïve participants tested four different spellers in a calibration and online task. All spellers were similar except the employed illumination of the target stimulus: white letters, white blocks, coloured letters, and coloured blocks.

**Results:** The block-shaped conditions offered an improvement versus the letter-shaped conditions in the calibration (*accuracy*) and online (*accuracy* and *correct commands per minute*) tasks. Analysis of the P300 waveform showed a larger difference between target and no target stimulus waveforms for the block-shaped conditions versus the letter-shaped. The hypothesis regarding the colour heterogeneity of the stimuli was not found at any level of the analysis.

**Conclusion:** The use of block-shaped illumination demonstrated a better performance than the standard letter-shaped flashing stimuli in classification performance, correct commands per minute, and P300 waveform.

## 1. Introduction

Brain-computer interfaces (BCI) are a technology that allow a user interact with the environment without the need for any kind of muscular activity (Wolpaw et al., 2002). Thus, this technology can offer a significant improvement in the quality of life of patients with severe motor reduction such as some lesions in the spinal cord or motor neuron diseases such as amyotrophic lateral sclerosis (ALS). The P300-based spellers could be considered the most widely studied devices since the publication by Farwell and Donchin (1988), who presented a virtual keyboard composed of a matrix of letters, which could be selected by the user to communicate.

The P300 is a positive deflection in the voltage of the electroencephalographic (EEG) signal, generally registered from the parietal lobe of the cortex, approximately 300 ms after the presentation of an uncommon target stimulus (Barrett, 1996). According to Nicolas-Alonso and Gomez-Gil (2012), the main advantages of P300-based systems are: i) they do not require extensive training for management; only a small calibration to adjust the system settings for each user; ii) they tend to have high success rates, and iii) they allow plenty of options to be chosen by the participant, due to the large number of stimuli that these systems allow using an oddball paradigm.

The classic speller presented by Farwell and Donchin (1988) consisted of a 6 × 6 matrix of letters and numbers, whose rows and columns were briefly intensified (i.e., flashed) a given number of times in a random order. The user should keep his/her attention over the target character and mentally count the number of times it was flashed. As this character was presented in one specific row and column, the P300 can be used to find the target stimulus using the oddball paradigm. Once a sequence of flashes was over, the symbol that belonged to the row and column that had produced the largest P300 was regarded as the attended character and given as feedback to the user.

Following the previously explained paradigm, numerous proposals have been suggested to improve the use of a P300 speller matrix (see Rezeika et al. (2018) for a detailed review). For example, some works have been focused on certain parameters such as variations in the lighting patterns (Townsend et al., 2010), presentation times and brightness intensity (Li et al., 2014), size of the stimuli and the distance between them (Li et al., 2011), colour (Takano et al., 2009), number of stimuli (Sellers et al., 2006), or the nature of these, e.g., letters, faces, or geometrical figures (Kaufmann et al., 2011; Treder et al., 2011).

Testing different lighting modalities of the stimuli that differ in the proportion of the illuminated stimulus surface could be convenient. An example of this lightning modality is the block-shaped illumination, i.e., illuminating the whole section around that character – for example, in a squared form – instead of illuminating the letter exclusively. Due to the increase in the illuminated surface size, the block-shaped paradigm could offer better results. Furthermore, one of the main advantages of the block-shaped illumination is that the size of the highlighted surface can be increased without the use of a larger surface of the graphical user interface (GUI), thus taking better advantage of the space. This advantage may be considerable in the case of mobile devices or those with a reduced screen size. Additionally, in the standard letter-shaped illumination, the size of the illuminated stimuli depends on the symbol size. Thus, the block-shaped illumination can homogenise the size between flashed stimuli.

---
[1] Author to whom any correspondence should be addressed.

The study of Speier et al. (2017) has been the only one that compared block-shaped illumination versus the standard condition of grey letters to white in a row-column paradigm. They showed a slight improvement over the classical condition in terms of *correct commands per minute* (*CCPM*) but not in *accuracy*; no further measures were carried out to assess the usability (e.g., analysis of calibration phase, EEG signal, or scores in subjective questionnaires). On the other hand, the use of different colours to illuminate the grey stimuli, instead of the classic grey to white illumination, has demonstrated an improvement in performance (Ryan et al., 2017, 2018). This demonstration was achieved using an 8 × 9 speller matrix with the so-called checkboard paradigm (Townsend et al., 2010), where the stimuli were flashed not in rows and columns but in groups with no adjacent flashing stimuli.

Therefore, the aim of the present study was to test these previous findings, regarding the hypothesis about the superiority of block-shaped and colour heterogeneity, combined in the same experiment. Combining these two factors in a single study – by carrying out a detailed analysis of the results – will allow us to study how both factors interact with each other. To do this, the present study will assess four speller matrixes which varied on the flashing stimuli: i) grey letters to white letters, ii) grey letters to coloured letters, iii) grey letters to white blocks, and iv) grey letters to coloured blocks. Statistical analyses were carried out to study the users' performance, event-related potential (ERP) waveform, and subjective questionnaire results. Based on the differences obtained between the different spellers it could be ascertained what factors – i.e., shape and/or colour heterogeneity of the flashing stimuli – could improve usability using a P300 BCI speller.

## 2. Methods

### 2.1. Participants

The study initially involved 15 participants (aged 22.67 ± 2.19, five males) who had normal or corrected-to-normal vision, identified as B1–B15. None of them had previous experience in BCI systems. Subjects were recruited through the use of social networks and posters around the campus. The study was approved by the Ethics Committee of the University of Malaga and met the ethical standards of the Helsinki Declaration. According to self-reports, none of the participants suffered recent (5 years) neurological or psychiatric illness or were taking any medication regularly. Participants received monetary remuneration of 10 € and all provided written informed consent.

### 2.2. Spelling paradigms

The present work used four different spellers that were handled and evaluated by the users. All were initially based on the previously mentioned row-column lighted paradigm of Farwell and Donchin (1988). Our proposals used a GUI, with a 3 × 4 matrix and a writing space at the top, whose visual angle was equal to 23.54º × 16.31º (25 cm × 17.2 cm, 60 cm away) displayed on a 15.6-in (39.6 cm) screen at a refresh rate of 60 Hz. The interface presented 12 targets: 10 letters, an underline, and a delete button to correct mistakes (denoted as "Bo"). In addition, a stimuli onset asynchrony (SOA) of 288 ms was used, and an interstimulus interval (ISI) of 96 ms, so each stimulus was presented for 192 ms. The pause between letters was equal to 5.47 s. The only differences between paradigms were the employed stimuli for each speller (Figure 1). Thus, the four presented paradigms were: i) white letters (WL), ii) coloured letters (CL), iii) white blocks (WB), and iv) coloured blocks (CB). All speller conditions are presented in Figure 1. The font used for the letters in all spellers was Arial bold in capital letters. Moreover, the size of the blocks was 4.49º × 3.34º of visual angle (4.7 cm × 3.5 cm).

### 2.3. Procedure

The experiment was carried out in an isolated room where only the participant was present at the time he/she was performing the task, so he/she could concentrate without external distractions. A within-subject design was used; thus, all users went through all the experimental conditions (i.e., the four spellers). Each user participated in two sessions in which two spellers were performed in each one. Due to experimental criteria, the time between sessions could not be less than 4 h or longer than 4 days. The order of the keyboard's presentation was selected pseudo-randomly following a Latin square design, so all were equally distributed to prevent any unwanted effects, such as learning of fatigue effects.

Each speller test consisted of two parts: i) an initial calibration task to adapt the system to the user and ii) an online writing test in which the user controlled the interface. Therefore, the main difference between both tasks was that in the first one the user did not have feedback, while in the second the aim of the user was to write the objective phrase.

Task 1: Calibration. In this phase, the specific user parameters were not yet available for the corresponding keyboard, so he/she could not obtain any feedback of their performance. The user had to "write" six words of five letters, a total of 30 characters with a short break between words (variable at the request of the user). The number of sequences (i.e., number of row-column flashes) was prefixed to 10, so each stimulus was lighted 20 times. The writing time for each character in this phase was equal to 20.06 s. Once these six words were "written", an offline analysis of the data was carried out to obtain the parameters that allowed him/her to continue with the session. The specific Spanish words were: "pares" (pairs), "chino" (Chinese), "presa" (dam), "hinco" (I sink), and "raspe" (scratch) y "nicho" (niche).

Task 2: Online writing test. This task consisted of writing the phrase "ceno_en_casa" (I have dinner at home, in Spanish), a sentence of 12 characters (10 letters and 2 underlines) using the different speller matrices. In this task, the number of sequences was adapted according to the obtained performance in the calibration task (i.e., task 1). The number of selected sequences was that in which the user obtained the second consecutive best accuracy. In cases where the maximum accuracy was not repeated consecutively or there was only one, the first best sequence was selected. Due to the feedback present in this task, the user could make mistakes in writing, so he/she had a command to delete the mistake and make a new attempt. The task ended once the user wrote the complete sentence.

At the end of each speller handling the user had to complete a questionnaire in relation to his/her experience during the control of the paradigm to assess usability of the system.

In addition, it should be avowed that participant B3 carried out the calibration task using the CL paradigm with 9 sequences instead 10, and B14 did not complete the subjective questionnaire for the WB speller. Also, the electrode P4 for E11 in his/her first session (WB and CB) and Fz for E12 in all spellers was removed due to artefact problems detected in the exploratory analysis.

### 2.4. Evaluation

For the calibration phase (task 1), the *accuracy* of letter selection and the waveform of the EEG signal were employed to assess performance. For the online test (task 2), the *accuracy*, the employed *time* to complete the task, and the *CCPM* were used (see (Speier et al., 2016) to obtain a more detailed view of the used metrics in the assessment of a speller matrix). Also, an *ad hoc* questionnaire to assess the user's experience during control of the device was used. This questionnaire included the following variables, ranging from 0 (very low) to 10 (very high): subjective perception of the performance (*performance*), difficulty to maintain attention on the interface (*attention*), mental effort required for controlling the speller (*mental effort*), level of comfort (*comfort*), level of frustration (*frustration*), and level of general satisfaction with the interface (*satisfaction*).

The present study shows four different types of spellers that have been compared. Therefore, for multiple comparison analysis, a correction method was applied because, as the number of conditions increased, the chance of getting a type I error (i.e., reject null hypothesis when it is true) is raised. Specifically, the *p* value limits rejection of the null hypothesis when equal to 0.05. Bonferroni correction was used in our study; however, this *p* value would be equivalent to 0.008 using the standard least significant difference test. In this way, despite losing some significant results that could be interesting, we can affirm that our conclusions are based on robust effects, barely due to randomness.

### 2.5. Data acquisition and signal processing

The EEG was recorded at a sample rate of 250 Hz using the electrode positions: Fz, Cz, Pz, Oz, P3, P4, PO7, and PO8, according to the 10/20 international system. All channels were referenced to TP8 and grounded to position FPz. Signals were amplified by an acti-CHamp amplifier (Brain Products GmbH, Munich, Germany). The amplifier settings were 0.5 and 1000 Hz for the band-pass filter, the notch filter (50 Hz) was on, and the sensitivity was 500 μV. All aspects of EEG data collection and processing were controlled by the BCI2000 system (Schalk et al., 2004). A stepwise linear discriminant analysis of the data was performed to obtain the weights for the P300 classifier and calculate the accuracy. Neither online nor offline artefact detection techniques were employed.

## 3. Results
### 3.1. Offline analysis
#### 3.1.1. Performance metrics

First, a three-way repeated measures analysis of variance (ANOVA) (2 × 2 × 10), including the factors of *colour* (white and coloured), *shape* (letters and blocks), and *sequence* (10 sequences) was carried out using *accuracy* as a dependent variable (Figure 2). The analysis showed a main effect of *shape* ($F(1, 12) = 8.102$; $p = 0.015$) and of *sequences* ($F(9, 108) = 46.171$; $p < 0.001$), but not in *colour* ($F(1, 12) = 1.089$; $p = 0.317$). The only interaction effect found was *shape*\**sequence* ($F(9, 108) = 11.686$; $p < 0.001$). Regarding the main effect of *shape*, as can be observed in Figure 2, block-shaped spellers (96.46 ± 1.31%) offered a higher performance than letter-shaped spellers (91.69 ± 2.07%) considering the sequences from 1 to 10. In addition, Figure 2 shows how the interaction effect denotes that, despite the fact that the learning curve was significantly better for block-shaped spellers than for letter-shaped spellers, the differences decreased as the sequences increased.

Additionally, a one-way repeated measure ANOVA, using the four spellers as factor and the chosen *number of sequences* as a dependent variable, showed that there were significant differences between spellers ($F(3, 42) = 3.842$; $p = 0.016$). The *number of sequences* needed to accomplish our criteria for each speller was equal to: 6.27 ± 2.37 for WL, 4.93 ± 1.91 for WB, 6.4 ± 2.13 for CL, and 4.73 ± 1.67 for CB. Specifically, there were statistical differences between CL and CB ($p = 0.018$), showing that the *number of sequences* was significantly inferior for CB than CL.

#### 3.1.2. P300 waveform

Figure 3 shows the grand average P300 waveform for target and non-target stimuli as a function of the four tested spellers. Next, analyses were performed using the eight previous mentioned electrodes in an interval time equal to 200–600 ms. At first, a three-way repeated measures ANOVA (2 × 2 × 8; *colour*, *shape*, and *channel*, respectively) was carried out. The *stimulus* factor (i.e., target versus no target stimuli) was not included because the dependent variable employed was the *amplitude difference* in absolute value between target and no target stimuli. This analysis showed two main effects: the *shape* ($F(1, 12) = 30.138$; $p < 0.001$) and the *channel* ($F(7, 84) = 5.781$; $p < 0.001$). Two interaction effects were also observed: *colour*\**shape* ($F(1, 12) = 8.21$; $p = 0.014$) and *colour*\**shape*\**channel* ($F(7, 84) = 2.26$; $p = 0.037$).

Regarding the *shape* main effect, the *amplitude difference* average for block-shaped was equal to 1.5 ± 0.44 µV, while the mean for letters was equal to 1.19 ± 0.39 µV. Thus, it could be admitted that the *amplitude difference* for block-shaped was higher than letter-shaped. According to *channel*, multiple comparisons showed that there were significant differences between next channels (the first indicated channel was significantly higher than the other): Fz (1.61 ± 0.7 µV) and P4 (0.77 ± 0.33 µV) ($p$ = 0.045); Cz (1.87 ± 0.72 µV) and Pz (1.29 ± 0.48 µV) ($p$ = 0.007); Cz (1.87 ± 0.72 µV) and P3 (1.04 ± 0.32 µV) ($p$ < 0.001); Cz and P4 (0.77 ± 0.33 µV) ($p$ < 0.001); Pz (1.29 ± 0.48 µV) and P4 (0.77 ± 0.33 µV) ($p$ < 0.001); and P3 (1.04 ± 0.32 µV) and P4 (0.77 ± 0.33 µV) ($p$ = 0.015). As can be observed Table 1, the electrodes Fz and Cz were those in which the greatest average values of *amplitude differences* were found. Likewise, due to the interaction effect of *colour*shape*, multiple comparisons between the four keyboards were performed: CB (1.63 ± 0.51 µV) and WL (1.22 ± 0.43 µV) ($p$ = 0.004); WB (1.37 ± 0.41 µV) and CL (1.16 ± 0.4 µV) ($p$ = 0.033); and CB (1.63 ± 0.51 µV) and CL (1.16 ± 0.4 µV) ($p$ < 0.001). Thus, the CB condition presented the greatest values in *amplitude differences*, offering significant results compared to the WL and CL conditions. Finally, in Table 1, the interaction effect of *colour*shape*channel* was studied comparing the differences between each keyboard individually for each electrode. Although significant differences between CB and WB were not observed, the CB condition offered the greatest *amplitude difference* values and, thus, the largest number of significant differences between conditions.

### 3.2. Online analysis

At first, three two-way ANOVA (2 × 2) with the factors *shape* and *colour* were carried out, one ANOVA for each employed variable in the online writing test: *accuracy*, *time* (time to write the phrase), and *CCPM* (Table 2). The only factor that showed significant differences was *shape* for *accuracy* ($F$ (1, 14) = 6.386; $p$ = 0.024), *time* ($F$ (1, 14) = 21.737; $p$ < 0.001), and *CCPM* ($F$ (1, 14) = 16.036; $p$ = 0.001). Specifically, the block-shaped spellers were better than letter-shaped in terms of *accuracy* (98.39 ± 3.42 and 94.94 ± 7.04, respectively), *time* (189.37 ± 48.98 and 254.76 ± 98.35, respectively), and *CCPM* (4.09 ± 0.93 and 3.33 ± 0.94, respectively). No interaction effect of *colour*shape* was observed in any variable.

### 3.3. Subjective questionnaires

In reference to the results obtained by the subjective questionnaire (Table 3), a two-way repeated measure ANOVA (2 × 2, *colour* and *shape*) was carried out for each item: *performance*, *attention*, *mental effort*, *comfort*, *frustration*, and *satisfaction*. No main effects were observed, so the *colour* and *shape* factors did not offer significant differences. However, there was an interaction effect between *colour* and *shape* (i.e., *colour*shape*) in two variables: *comfort* ($F$ (1, 13) = 6.443; $p$ = 0.025) and *satisfaction* ($F$ (1, 13) = 8.516; $p$ = 0.012). This interaction effect requires multiple comparisons between the four spellers. On the one hand, despite the previous interaction effect in *comfort*, there was no specific difference between spellers. On the other hand, for *satisfaction*, the differences were only obtained between CB (9.00 ± 1.30) and CL (7.29 ± 2.37) ($p$ = 0.049), where the CB condition offered the best results.

## 4. Discussion

We hypothesised that performance could be improved by applying an increased stimulus illuminated surface and/or a heterogeneous characteristic between stimuli (i.e., each adjacent stimulus is flashed in a different way). This proposal was carried out using the block-shaped condition and colour heterogeneity. It was tested at many levels: in the calibration task through *accuracy* and the study of the P300 waveform, in the online writing task through *accuracy*, *time*, and *CCPM*, and with the subjective questionnaire.

In the calibration task, as expected, most participants reached an accuracy close to 100% at some point in the 10 sequences, since the criteria to choose the *number of sequences* ranged between the fourth and seventh sequence (6.27 ± 2.37 for WL, 4.93 ± 1.91 for WB, 6.4 ± 2.13 for CL, and 4.73 ± 1.67 for CB). In addition, all spellers equalise the *accuracy* in the last sequences. Therefore, the goal was not only to reach that maximum, but doing so with as few sequences as possible. According to our results, the block-shaped spellers achieved this aim. An example to illustrate this finding is that most participants overcame 90% accuracy with only two sequences using the block-shaped, while for the letter-shape spellers it required four sequences (Figure 2).

The P300 waveform analysis showed that the block-shaped spellers offered higher differences between the target and no target stimulus EEG signal (i.e., *amplitude difference*), suggesting that the processing carried out by the user could be modified by the shape of the stimulus. If the four spellers are considered individually, the CB speller (1.63 ± 0.51 µV) could be suggested as that in which the larger significant *amplitude difference* was found versus WL (1.22 ± 0.43 µV) and CL (1.16 ± 0.4 µV), while the WB (1.37 ± 0.41 µV) only offered significant differences versus the CL (1.16 ± 0.4 µV). In addition, the channel Cz, followed by Fz, was shown as the electrode that offered the largest *amplitude differences*, offering significant differences versus electrodes from the parietal zone exclusively (i.e., Pz, P3, and P4). Despite using a different dependent variable, these results are consistent with a previous study (Jarmolowska et al., 2013), in which the highest values in µV for the target stimulus included the electrodes Fz and Cz. In addition, Ryan et al., (2017) showed that the central electrodes (grouped as Fz, Cz, and Pz) gave rise to a greater negativity and positivity (in a time window of 113–515 ms) for the target stimulus than those located in the parieto-occipital zones (grouped as P3, P4, PO7, PO8, and Oz).

The results in the online task were similar to those obtained in the calibration, in which a superiority of the block-shaped paradigms was shown. Thus, these results reinforced the conclusions obtained in the calibration phase. In addition, it worth noting that, although the *number of sequences* was chosen following the same criteria for all spellers, there were significant differences in *shape* factor for *accuracy* (98.39 ± 3.42 for block-shaped and 94.94 ± 7.04 for letter-shaped). Therefore, the worst performance in the online task for letter-shaped was not due to the higher *number of sequences*, but due to the speller itself. Despite the study of Speier et al. (2017) offered non-significant differences for accuracy in the online writing test, these differences were found in the present study.

Regarding the *time* for writing the phrase and the *CCPM*, similar results to Speier et al. (2017) were found in the present study: the block-shaped conditions (189.37 ± 48.98 s and 4.09 ± 0.93 of *CCPM*) offered significant differences versus the letter-shaped ones (254.76 ± 98.35 s and 3.33 ± 0.94 of *CCPM*).

In reference to the subjective questionnaires, only the *satisfaction* variable showed significant differences between the CB and CL keyboard, with the CB speller being the most satisfying keyboard. Despite not having obtained significant differences in the rest of the variables, it should be noted that the CB speller obtained better scores than the rest in all the variables recorded by the questionnaire. Therefore, we suggest that the CB speller was the most preferred by the participants. Similarly, the CL keyboard showed the worst scores in each of the variables, thus we suggest that it was the least preferred by the participants. The low performance of the CL keyboard could be due to the difficulty to perceive the illuminations, since 6 of the 9 users commented that some colours were difficult to perceive due to the low contrast with respect to the background.

Regarding the colour heterogeneity hypothesis, an unexpected result of our study is that it did not replicate the results obtained by Ryan et al. (2017). While Ryan et al. (2017) showed that the use of different colours for the illumination of the letters produced an improvement of the performance, the present study obtained that precisely the condition of the coloured-letters (i.e., CL) was the speller with the lowest performance since it showed the greater number of significant differences obtaining the worst scores. At the moment, we are not able to offer an adequate explanation for our results besides the differences between both experiments. On the one hand, the size of the matrix was considerably different, while our matrix was 3 × 4, the matrix used by Ryan et al. (2017) was 8 × 9. Also, the paradigm used by them was not the row-column, but the checkboard, as proposed by Townsend et al. (2010), in which the elements were illuminated by sets previously established instead of rows and columns, so that the flash of the target stimulus did not temporarily coincide with the adjacent stimuli.

Finally, we would like to recall that use of the Bonferroni correction could lead to the presence of type II errors (i.e., accept null hypothesis when it is false). Thus, some significant differences could have been missed. On the contrary, it should be affirmed that the significant results obtained in the present study show strong effects, which are rarely observed by randomness.

## 5. Conclusions

The present study has demonstrated a clear superiority of the use of the block-shaped condition for the illumination of the stimulus versus the exclusive illumination of the letter. Therefore, the previous results obtained by Speier et al. (2017), which needed more extensive tests – such as a report of the performance in calibration task and ERP analysis – to verify the finding related to the use of block-shaped, has been corroborated. Specifically, the present work obtained significant differences regarding *shape* factor in both phases of the study: calibration task (*accuracy* and ERP waveform) and online task (*accuracy*, *time* and *CCPM*). Additionally, although specific differences between the two block-shaped spellers (i.e., WB and CB) were not found, we suggest that – based on the questionnaire results – the CB keyboard was preferable. The CB condition offered the greatest number of significant differences in this study, such as the *accuracy* obtained in the calibration phase or the satisfaction in the subjective questionnaire.

In future experiments, it would be interesting to test the block-shaped paradigm by varying the number of elements of the matrix since – although it is possible to implement a keyboard with 12 stimuli that allows a complete writing (e.g., Ron-Angevin et al. (2015)) – the most common is a 6 × 6 instead of a 3 × 4 matrix. Also, although the block-shaped offers a larger illuminated stimulus surface than solely illumination of the letter, the block-shaped illumination can use the space more efficiently in small screens. Thus, testing different matrix sizes using the block-shaped paradigm should be studied. Finally, considering that patients with motor diseases, such as ALS, are the main objective of these devices, it would be convenient to reproduce these results in that population.

## Acknowledgements

This work was partially supported by the Spanish Ministry of Economy and Competitiveness through the project LICOM (DPI2015-67064-R), by the European Regional Development Fund (ERDF) and by the University of Malaga. Moreover, the authors would like to thank all participants for their cooperation.

**Figures and tables**

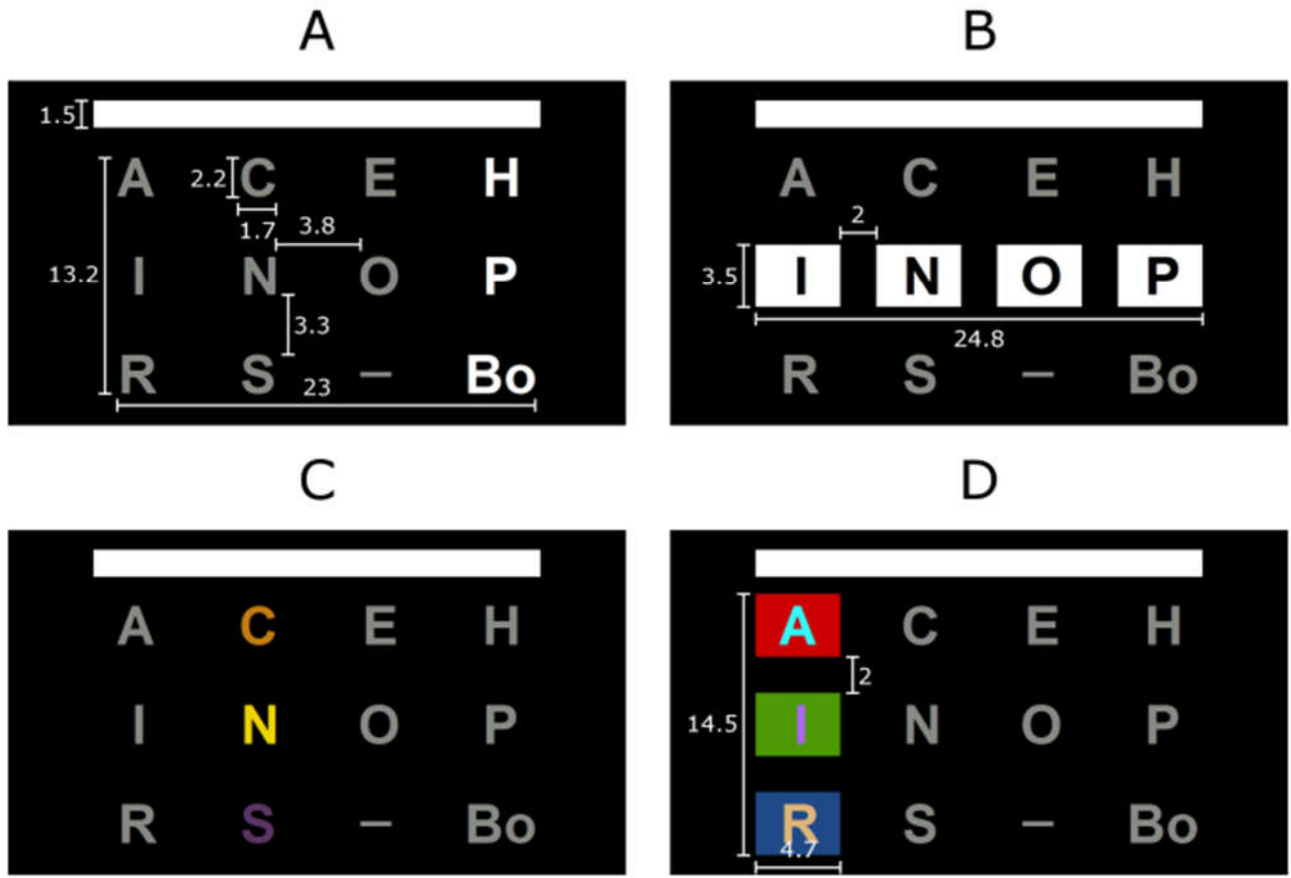

Figure 1. The four conditions employed in the experiment. A) White letters (WL), B) white blocks (WB), C) coloured letters (CL), and D) coloured blocks (CB). Distances are represented in centimetres.

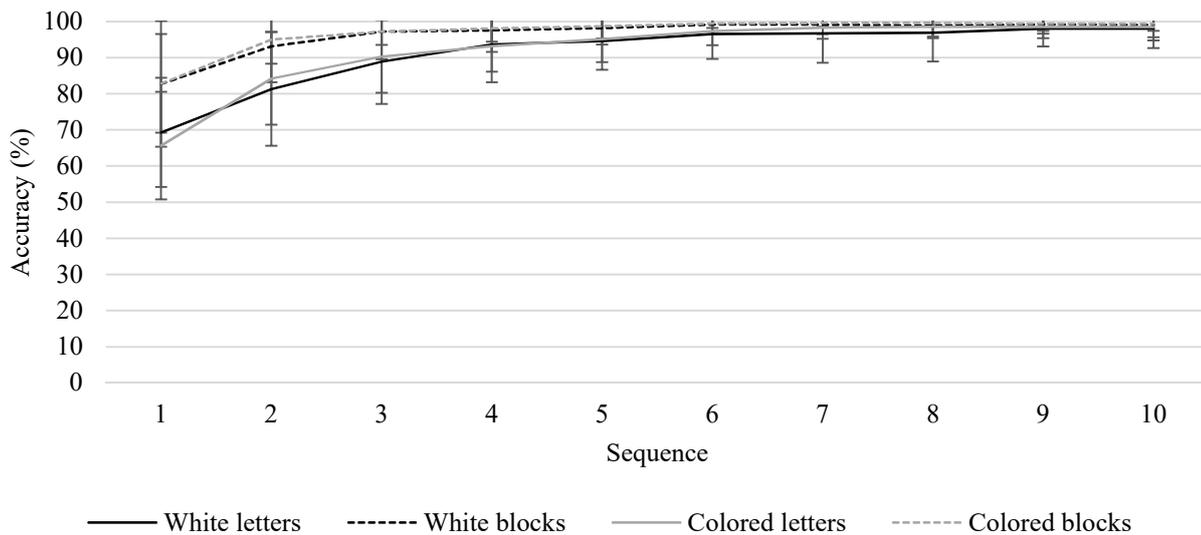

Figure 2. Average accuracies (%, mean ± standard deviation) of the different P300-speller conditions as a function of the number of sequences in the calibration task.

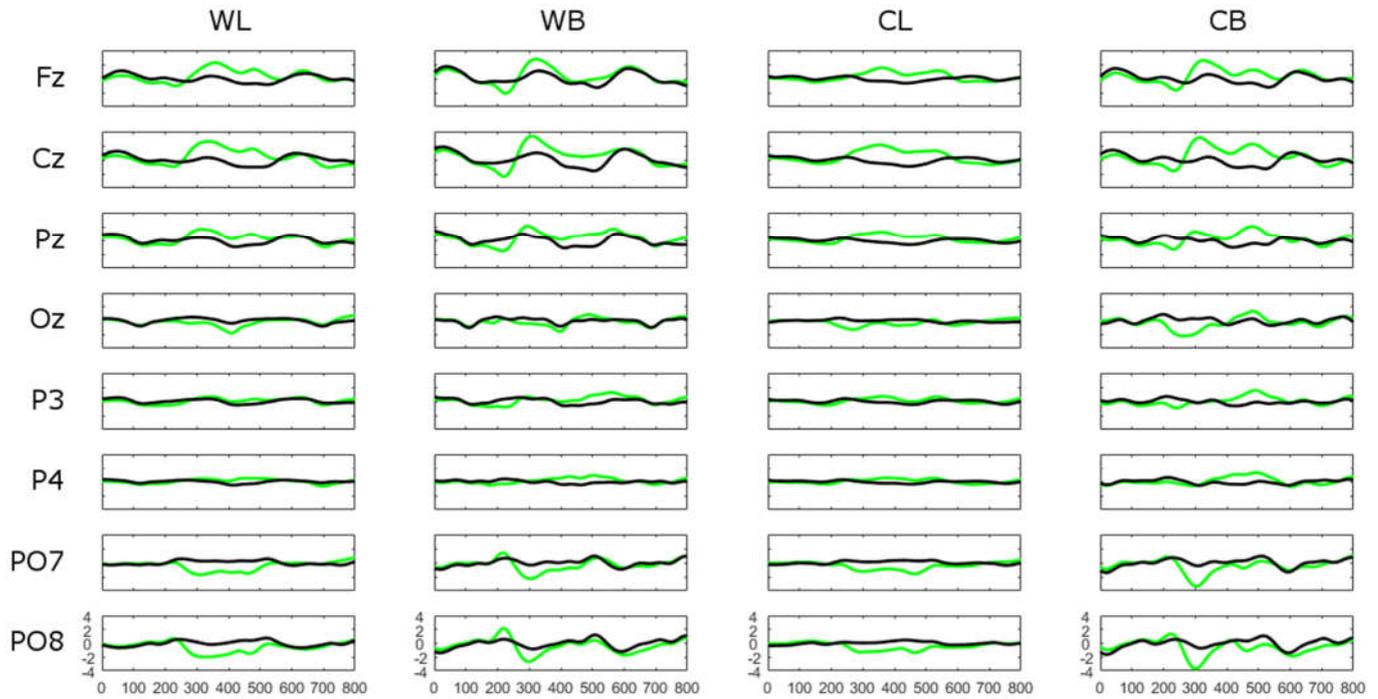

Figure 3. Grand averaged P300 waveforms for target (green) and non-target stimuli (black) for the eight used electrodes and keyboards (white letters (WL), white blocks (WB), coloured letters (CL) and coloured blocks (CB)).

TABLE 1. Amplitude differences (μV, mean ± standard deviation) between target and no target stimuli in absolute value for each channel.

| Channel | Speller | | | | Average |
|---|---|---|---|---|---|
| | (1) WL | (2) WB | (3) CL | (4) CB | |
| Fz | 1.6 ± 0.79 | 1.52 ± 0.54 | 1.35 ± 0.8 [4] | 1.98 ± 0.96 [3] | 1.61 ± 0.7 |
| Cz | 1.71 ± 0.56 | 1.82 ± 0.82 | 1.77 ± 0.84 [4] | 2.17 ± 0.96 [3] | 1.87 ± 0.72 |
| Pz | 1.05 ± 0.48 [4] | 1.42 ± 0.57 | 1.12 ± 0.53 [4] | 1.56 ± 0.61 [1,3] | 1.29 ± 0.48 |
| Oz | 1.09 ± 0.69 [4] | 1.26 ± 0.59 | 1.04 ± 0.4 [4] | 1.62 ± 0.75 [1,3] | 1.25 ± 0.53 |
| P3 | 0.87 ± 0.39 [4] | 1.17 ± 0.43 | 0.85 ± 0.36 [4] | 1.27 ± 0.5 [1,3] | 1.04 ± 0.32 |
| P4 | 0.59 ± 0.26 [2,4] | 0.91 ± 0.37 [1,3] | 0.64 ± 0.35 [2,4] | 1.01 ± 0.49 [1,3] | 0.77 ± 0.33 |
| PO7 | 1.48 ± 1.24 | 1.6 ± 1.02 | 1.37 ± 0.95 [4] | 1.81 ± 1.07 [3] | 1.56 ± 1.02 |
| PO8 | 1.4 ± 0.78 | 1.24 ± 0.53 [4] | 1.11 ± 0.58 [4] | 1.59 ± 0.69 [2,3] | 1.34 ± 0.61 |
| Average | 1.22 ± 0.43 [4] | 1.37 ± 0.41 [3] | 1.16 ± 0.4 [2,4] | 1.63 ± 0.51 [1,3] | |

Note: White letters (WL), coloured letters (CL), white blocks (WB), and coloured blocks (CB). Significant differences between spellers ($p < 0.05$) are denoted with a superindex to show which spellers the average was different to (1 for WL, 2 for WB, 3 for CL, and 4 for CB). The Bonferroni correction was applied.

TABLE 2. Accuracy, time and correct commands per minute (*CCPM*) (mean ± standard deviation) in the online writing test for the different P300-speller conditions.

| Variable | Speller | | | |
|---|---|---|---|---|
| | (1) WL | (2) WB | (3) CL | (4) CB |
| Accuracy (%) | 95.6 ± 5.22 | 98.21 ± 4.88 | 94.27 ± 8.86 | 98.57 ± 2.96 |
| Time (s) | 243.67 ± 84.91 | 193.26 ± 55.07 [3] | 265.83 ± 111.79 [2,4] | 185.48 ± 42.87 [3] |
| CCPM | 3.40 ± 0.96 | 4.03 ± 0.87 [3] | 3.26 ± 0.92 [2,4] | 4.15 ± 0.98 [3] |

Note: White letters (WL), coloured letters (CL), white blocks (WB), and coloured blocks (CB). Significant differences between spellers ($p < 0.05$) are denoted with a superindex to show which spellers the average was different to (1 for WL, 2 for WB, 3 for CL, and 4 for CB). The Bonferroni correction was applied for multiple comparisons.

TABLE 3. Scores (mean ± standard deviation) for the variables collected in the subjective questionnaire for the different P300-speller conditions.

| Variable | Speller | | | |
|---|---|---|---|---|
| | (1) WL | (2) WB | (3) CL | (4) CB |
| Performance | 8.29 ± 1.07 | 8.43 ± 1.34 | 7.71 ± 1.73 | 8.57 ± 1.22 |
| Attention | 7.57 ± 2.47 | 7.86 ± 2.11 | 7.07 ± 2.34 | 8.36 ± 1.86 |
| Mental effort | 5.07 ± 2.87 | 4.79 ± 3.04 | 5.71 ± 2.76 | 4.36 ± 2.95 |
| Comfort | 8.07 ± 1.77 | 7.57 ± 2.31 | 6.71 ± 2.97 | 8.36 ± 1.98 |
| Frustration | 2.29 ± 2.76 | 2.79 ± 3.02 | 3.57 ± 2.98 | 1.93 ± 2.37 |
| Satisfaction | 8.86 ± 1.41 | 8.71 ± 1.49 | 7.29 ± 2.37 [4] | 9.00 ± 1.30 [3] |

Note: White letters (WL), coloured letters (CL), white blocks (WB), and coloured blocks (CB). Significant differences between spellers ($p < 0.05$) are denoted with a superindex to show which spellers the average was different to (1 for WL, 2 for WB, 3 for CL, and 4 for CB). The Bonferroni correction was applied for multiple comparisons.